\documentclass[conference]{IEEEtran}
\IEEEoverridecommandlockouts

\usepackage{cite}
\usepackage{stfloats}
\usepackage{tikz}
\usepackage{wrapfig,bm}
\usetikzlibrary{arrows.meta, positioning, calc, positioning, shapes.geometric}

\newdimen\arrowsize

\pgfarrowsdeclare{smallstealth}{smallstealth}
{
  \arrowsize=2pt 
  \advance\arrowsize by .5\pgflinewidth
}
{
  \pgfsetdash{}{0pt}
  \pgfsetroundjoin
  \pgfsetmiterjoin
  \pgfsetlinewidth{\pgflinewidth}
  \pgfpathmoveto{\pgfpoint{0pt}{0pt}}
  \pgfpathlineto{\pgfpoint{-2pt}{1pt}} 
  \pgfpathlineto{\pgfpoint{-1.5pt}{0pt}} 
  \pgfpathlineto{\pgfpoint{-2pt}{-1pt}}
  \pgfpathclose
  \pgfusepathqfillstroke
}

\usepackage{xcolor}
\usepackage{amsmath,amssymb,amsfonts}
\usepackage{algorithm}
\usepackage{amsthm}
\usepackage{algorithmic}
\usepackage{graphicx}
\usepackage{textcomp}
\usepackage[left=0.625in,right=0.625in,top=0.75in,bottom=1in]{geometry}
\def\BibTeX{{\rm B\kern-.05em{\sc i\kern-.025em b}\kern-.08em
    T\kern-.1667em\lower.7ex\hbox{E}\kern-.125emX}}

\newcommand{\ignore}[1]{}
\usepackage{color}

\begin{document}



\title{Edge Server Monitoring for Job Assignment \\



\thanks{This work has been supported in part by the Army Research Laboratory and was accomplished under Cooperative Agreement Number W911NF-24-2-0205, and by the U.S. National Science Foundation under the grant CNS-2239677. The views and conclusions contained in this document are those of the authors and should not be interpreted as representing the official policies, either expressed or implied, of the Army Research Laboratory or the U.S. Government. The U.S. Government is authorized to reproduce and distribute reprints for Government purposes notwithstanding any copyright notation herein.}
}

\author{
\IEEEauthorblockN{Samuel Chamoun, Sirin Chakraborty}
\IEEEauthorblockA{\textit{Department of ECE,} 
\textit{Auburn University}\\
Auburn, AL, USA 
\\
\{sgc0042,szc0260\}@auburn.edu
}
\and
\IEEEauthorblockN{Eric Graves, Kevin Chan}
\IEEEauthorblockA{\textit{DEVCOM}
\textit{Army Research Laboratory} \\
Adelphi, MD, USA \\
\{eric.s.graves9.civ,kevin.s.chan.civ\}@army.mil
}
\and
\IEEEauthorblockN{Yin Sun}
\IEEEauthorblockA{\textit{Department of ECE,} 
\textit{Auburn University}\\
Auburn, AL, USA 
\\
yzs0078@auburn.edu
}
}

\maketitle

\begin{abstract}
In this paper, we study a goal-oriented communication problem for edge server monitoring, where compute jobs arrive intermittently at dispatchers and must be immediately assigned to distributed edge servers. Due to competing workloads and the dynamic nature of the edge environment, server availability fluctuates over time. To maintain accurate estimates of server availability states, each dispatcher updates its belief using two mechanisms: (i) active queries over shared communication channels and (ii) feedback from past job executions. We formulate a query scheduling problem that maximizes the job success rate under limited communication resources for queries. This problem is modeled as a Restless Multi-Armed Bandit (RMAB) with multiple actions and addressed using a Net-Gain Maximization (NGM) scheduling algorithm, which selects servers to query based on their expected improvement in execution performance. Simulation results show that the proposed NGM Policy significantly outperforms baseline strategies, achieving up to a 30\% gain over the Round-Robin Policy and up to a 107\% gain over the Never-Query Policy.
\end{abstract}

\section{Introduction}
In edge computing systems deployed in contested or resource-constrained environments, maintaining a high success rate in compute job execution is critical for supporting mission-critical, real-time edge services, such as Intelligence, Surveillance, and Reconnaissance (ISR), enhanced situational awareness in dynamic operational theaters, and coordinated control of autonomous systems in multi-agent missions. For example, a tactical surveillance system may need to process live video streams from multiple ground or aerial sensors, requiring dispatchers to rapidly assign computationally intensive jobs to edge servers. However, server availability in such environments is often uncertain and highly dynamic, influenced by factors such as mobility, wireless interference, and competing workloads. To assess server availability under these conditions, conventional systems typically rely on periodic polling to retrieve status updates from edge servers. These approaches, however, introduce significant communication overhead and scale poorly in bandwidth-constrained environments, ultimately limiting the system’s ability to maintain high job success rates under strict query budgets.

In this paper, we study a goal-oriented communication problem for edge server monitoring, where compute jobs arrive intermittently at dispatchers and must be immediately assigned to one of their servers. Each dispatcher maintains estimates of the idle or busy status of its servers, which are updated through two mechanisms: (i) active queries over shared communication channels and (ii) feedback from past job executions. 
The servers are partitioned into non-overlapping groups, where each dispatcher manages and assigns jobs exclusively to the servers within its group. In each time slot, a dispatcher may issue a query to one of its servers to directly observe its current status, thereby reducing uncertainty in its estimate.  Due to limited communication resources, all dispatchers share a common pool of communication channels, which constrains the total number of queries that can be issued system-wide in each time slot. Additionally, when a job is assigned to a server, the observed outcome, i.e., success if the server is idle, or failure if it is busy, provides implicit feedback that further refines the dispatcher’s belief. Each job completes execution within a single time slot. The technical contributions of this paper are summarized as follows:
\begin{itemize}
    \item We express the job success rate as a function of the information vector, which includes the most recently observed server status and the Age of Information (AoI). The AoI measures the freshness of the most recently observed server status, updated through queries and implicit feedback. Leveraging this relationship, we formulate a query scheduling problem that aims to maximize the long-term average job success rate, subject to per-time-slot query budget constraints.
    
    \item We model the query scheduling problem as a Restless Multi-Armed Bandit (RMAB) with multiple actions and address it using a Net-Gain Maximization (NGM) algorithm \cite{shisher2023learning,ornee2023context, chakraborty2024timely}. The algorithm selects servers to query based on their net-gains, i.e., the expected improvement in job execution performance, and selects servers with the highest non-negative gains. 
    \item Numerical results show that the proposed NGM Policy significantly outperforms baseline strategies, achieving up to a 30\% gain over the Round-Robin Policy and up to a 107\% gain over the Never-Query Policy. 
\end{itemize}


\section{Related Work}

The Age of Information (AoI) has become a popular metric for quantifying data freshness in real-time systems ~\cite{kaul2012real}. Early works established AoI as a fundamental metric for data freshness in real-time systems, focusing primarily on queue management and scheduling strategies to minimize average AoI \cite{kaul2012real, sun2016update, sun2017update, kadota2018optimizing}. Subsequent works incorporated information-theoretic insights, demonstrating that metrics such as mutual information and feature–time correlation can more accurately quantify the utility of each update \cite{sun2018information, sun2019sampling}. In parallel, the AoI framework was extended to remote estimation settings to reveal direct connections between stale information and estimation error \cite{sun2017remote, chakraborty2024timely, tsai2021unifying}. Building on these, purely information‐theoretic analyses in both Markovian settings \cite{sun2017remote} and autoregressive (non‐Markovian) models \cite{shisher2021age, shisher2024monotonicity} have precisely characterized how delays degrade estimate timeliness, and simulations confirm that stale features substantially impair prediction accuracy in remote inference \cite{shisher2023learning}.


More recently, several studies have applied AoI principles to goal‐oriented communications. For example, in remote inference systems, AoI‐aware sampling and transmission policies have been shown to improve prediction accuracy by prioritizing timely feature updates \cite{shisher2022does, shisher2023learning, shisher2024timely}. AoI has also been leveraged for multimodal edge inference \cite{zhang2025mass} and for remote safety monitoring to keep critical state information fresh \cite{ornee2023context, ornee2023remote}; a related line studies significance-aware status updating to prioritize updates by situational relevance \cite{ornee2025remote}. Alongside these developments, the concept of Age of Channel State Information (AoCSI) was introduced to guide decisions about pilot versus data transmission, striking a balance between acquiring up‐to‐date channel estimates and maximizing throughput \cite{costa2015, chakraborty2025send}. Complementary to these works, joint computation and communication co-scheduling for timely multi-task inference at the wireless edge has been studied in~\cite{shisher2025infocom}, which emphasizes coordinated resource allocation for inference accuracy.


To that end, recent studies have focused on optimizing update policies for distributed monitoring jobs to improve data freshness in real-time sensing systems~\cite{graves2024optimal}, typically assuming a single-agent update model. Parallel work on freshness-aware scheduling for dynamic job assignment in Markovian environments adapts allocation to the expected system state but considers single job-to-server settings~\cite{banerjee2025tracking}. 
More recently, \cite{chamoun2025milcom} examined cooperative multi-agent edge-server monitoring, using a MAPPO-based policy to co-optimize querying and dispatch. However, this approach relies on centralized training and coordination, and prior works still do not address the fully decentralized setting where multiple dispatchers compete for limited communication resources while jointly leveraging explicit queries and implicit execution feedback.

\begin{figure}[t]
    \centering
    \scalebox{0.925}{\begin{tikzpicture}[
    >=Stealth, 
    auto, 
    node distance=1.2cm and 1.5cm, 
    scale=0.75, 
    transform shape,
    rect/.style={
        draw, 
        rectangle,  
        minimum width=2.4cm, 
        minimum height=0.75cm,
        align=center},
    rectChannel/.style={
        draw, 
        rectangle, 
        minimum width=2.2cm, 
        minimum height=0.75cm,
        align=center},
    rectServer/.style={
        draw, 
        rectangle, 
        minimum width=3cm, 
        minimum height=0.75cm,
        align=center}
]

\node[rect] (D1) {\textbf{Dispatcher 1}};
\node[rect, below=0.5cm of D1] (D2) {\textbf{Dispatcher 2}};
\node[draw=none, below=0.25cm of D2] (dots1) {\large\textbf{$\vdots$}};
\node[rect, below=0.5cm of dots1] (DN) {\textbf{Dispatcher} $\boldsymbol{N}$};

\node[draw=none, right=1.05cm of dots1] (dots2) {\large\textbf{$\vdots$}};

\node[rectChannel, right=1.75cm of D1] (C1) {\textbf{Channel 1}};
\node[rectChannel, right=1.75cm of D2] (C2) {\textbf{Channel 2}};
\node[draw=none, below=0.25cm of C2] (dots3) {\large\textbf{$\vdots$}};
\node[draw=none, left=1cm of dots3] (dots7) {\large\textbf{$\vdots$}};
\node[rectChannel, right=1.75cm of DN] (CM) {\textbf{Channel} $\boldsymbol{M}$};

\node[draw=none, right=1cm of dots3] (dots4) {\large\textbf{$\vdots$}};
\node[rectServer, right=1.75cm of C1] (S11) {\textbf{Server} $\boldsymbol{X_{1,1}(t)}$};
\node[rectServer, right=1.75cm of C2] (S12) {\textbf{Server} $\boldsymbol{X_{1,2}(t)}$};
\node[draw=none, below=0.25cm of S12] (dots5) {\large\textbf{$\vdots$}};
\node[draw=none, left=1.4cm of dots5] (dots7) {\large\textbf{$\vdots$}};
\node[rectServer, right=1.75cm of CM] (SMK) {\textbf{Server} $\boldsymbol{X_{N,K}(t)}$};


\node[left=0.5cm of D1] (AID1) {\textbf{Jobs}};
\node[left=0.5cm of D2] (AID2) {\textbf{Jobs}};
\node[draw=none, below=0.25cm of AID2] (dots6) {\large\textbf{$\vdots$}};
\node[left=0.5cm of DN] (AIDN) {\textbf{Jobs}};

\draw[->, >=smallstealth, line width=0.5pt] (AID1.east) -- (D1.west);
\draw[->, >=smallstealth, line width=0.5pt] (AID2.east) -- (D2.west);
\draw[->, >=smallstealth, line width=0.5pt] (AIDN.east) -- (DN.west);

\draw[<->, >=smallstealth, line width=0.5pt] ($(D1.east)+(0,6pt)$) -- ($(C1.west)+(0,6pt)$);
\draw[<->, >=smallstealth, line width=0.5pt] ($(D1.east)+(0,0pt)$) -- ($(C2.west)+(0,6pt)$);
\draw[<->, >=smallstealth, line width=0.5pt] ($(D1.east)+(0,-6pt)$) -- ($(CM.west)+(0,6pt)$);

\draw[<->, >=smallstealth, line width=0.5pt] ($(D2.east)+(0,6pt)$) -- ($(C1.west)+(0,0pt)$);
\draw[<->, >=smallstealth, line width=0.5pt] ($(D2.east)+(0,0pt)$) -- ($(C2.west)+(0,0pt)$);
\draw[<->, >=smallstealth, line width=0.5pt] ($(D2.east)+(0,-6pt)$) -- ($(CM.west)+(0,0pt)$);

\draw[<->, >=smallstealth, line width=0.5pt] ($(DN.east)+(0,6pt)$) -- ($(C1.west)+(0,-6pt)$);
\draw[<->, >=smallstealth, line width=0.5pt] ($(DN.east)+(0,0pt)$) -- ($(C2.west)+(0,-6pt)$);
\draw[<->, >=smallstealth, line width=0.5pt] ($(DN.east)+(0,-6pt)$) -- ($(CM.west)+(0,-6pt)$);

\draw[<->, >=smallstealth, line width=0.5pt] ($(C1.east)+(0,6pt)$) -- ($(S11.west)+(0,6pt)$);
\draw[<->, >=smallstealth, line width=0.5pt] ($(C1.east)+(0,0pt)$) -- ($(S12.west)+(0,6pt)$);
\draw[<->, >=smallstealth, line width=0.5pt] ($(C1.east)+(0,-6pt)$) -- ($(SMK.west)+(0,6pt)$);

\draw[<->, >=smallstealth, line width=0.5pt] ($(C2.east)+(0,6pt)$) -- ($(S11.west)+(0,0pt)$);
\draw[<->, >=smallstealth, line width=0.5pt] ($(C2.east)+(0,0pt)$) -- ($(S12.west)+(0,0pt)$);
\draw[<->, >=smallstealth, line width=0.5pt] ($(C2.east)+(0,-6pt)$) -- ($(SMK.west)+(0,0pt)$);

\draw[<->, >=smallstealth, line width=0.5pt] ($(CM.east)+(0,6pt)$) -- ($(S11.west)+(0,-6pt)$);
\draw[<->, >=smallstealth, line width=0.5pt] ($(CM.east)+(0,0pt)$) -- ($(S12.west)+(0,-6pt)$);
\draw[<->, >=smallstealth, line width=0.5pt] ($(CM.east)+(0,-6pt)$) -- ($(SMK.west)+(0,-6pt)$);

\end{tikzpicture}}  
    \caption{System model: Jobs arrive intermittently at each of the $N$ dispatchers. Upon arrival, a job is immediately assigned by the dispatcher to one of its $K$ servers. Each dispatcher estimates the availability of its $K$ servers using two sources of information: (i) active queries transmitted over $M$ shared communication channels and (ii) feedback obtained from past job executions.}
    \label{fig:system_model}
\end{figure}
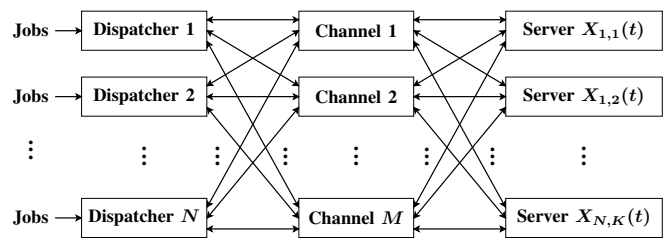

\section{Model and Formulation}
\subsection{System Model}
Consider an edge AI system that includes \(N\) job dispatchers, $M$ communication channels ($M< N$), and \(NK\) edge servers, as illustrated in Fig.~\ref{fig:system_model}. The edge servers are partitioned into $N$ non-overlapping groups, each consisting of $K$ servers. 
To facilitate job assignment upon job arrivals, each dispatcher~\(n\) estimates the idle/busy status of the \(K\) servers in its group at each time slot \(t \in \{0,1,\dots\}\). Let $X_{n,k}(t) \in \{0,1\}$ represent the idle/busy status of the $k$-th server managed by dispatcher \(n\), where
\begin{IEEEeqnarray}{rCl}
   \!\!\!\!\!X_{n,k}(t)\! =\! 
   \begin{cases}
       0, & \!\!\!\!\!\text{if server \(k\) of dispatcher $n$ is idle at time } t,\\
       1, & \!\!\!\!\!\text{if server \(k\) of dispatcher $n$ is busy at time } t.
   \end{cases}
   \label{eq:server_status}
\end{IEEEeqnarray}
Due to other background computation jobs on edge servers, $X_{n,k}(t)$ varies over time. To model this time-dependent behavior, we assume that \(X_{n,k}(t)\) evolves as an ergodic binary Markov chain with transition matrix 
\begin{IEEEeqnarray}{rCl}
   \bm P_{n,k} =
   \begin{bmatrix}
       \phi_{n,k} & 1 - \phi_{n,k} \\
       1 - \psi_{n,k} & \psi_{n,k}
   \end{bmatrix},
   \label{eq:transition_matrix}
\end{IEEEeqnarray}
where \(\phi_{n,k} \in (0,1)\) represents the probability that server \(k\), if idle at time slot \(t\), remains idle at time slot \(t+1\); and \( \psi_{n,k}\in(0,1)\)  is the probability that server \(k\), if busy at time slot \(t\), remains busy at time slot \(t+1\). Let $\boldsymbol{\pi}_{n,k} = \begin{bmatrix} \pi_{n,k}^{(0)} & \pi_{n,k}^{(1)} \end{bmatrix}$ denote the stationary distribution of the Markov chain $X_{n,k}(t)$, which can be obtained by solving the following equations:
\begin{IEEEeqnarray}{rCl}
\boldsymbol{\pi}_{n,k} \bm P_{n,k} = \boldsymbol{\pi}_{n,k}, \quad \pi_{n,k}^{(0)} + \pi_{n,k}^{(1)} = 1.
\label{eq:stationary_distribution}
\end{IEEEeqnarray}
The system starts to operate at time slot $t = 0$, with $X_{n,k}(0)$ assumed to follow the stationary distribution $\boldsymbol{\pi}_{n,k}$.

The arrivals of job requests are modeled as a Bernoulli process. Let \(z_n(t) \in \{0,1\}\) denote a Bernoulli random variable such that \(z_n(t) = 1\) if a job arrives at dispatcher \(n\) at time \(t\), and \(z_n(t) = 0\) otherwise. The random variables \(z_n(t)\) are independent across dispatchers and \emph{i.i.d.} across time slots, with mean  \(\mathbb E[z_n(t)] = \lambda_n\). 
If a job arrives at time $t$ (i.e., $z_n(t) = 1$), dispatcher~$n$ immediately assigns it to one of its $K$ servers. If the selected server is available, the job is successfully completed within a single time slot; otherwise, it is dropped. 
We leave the consideration of job buffering and failover to future work, but note that the Net-Gain Maximization scheduling policy in Section \ref{sec:rmab} is sufficiently powerful to handle these more general cases effectively. 

\subsection{Job Success Rate}
Dispatchers update their estimates of server availability using both active queries and acknowledgments (both positive ACKs and negative NAKs) from servers after job execution. Let \(c_{n,k}(t) \in \{0,1\}\) denote whether dispatcher $n$ queries its server $k$ at time $t$, defined as
\begin{IEEEeqnarray}{rCl}
\!\!\!\!\!\!c_{n,k}(t) \!=\!
\begin{cases}
1, &\!\!\!\! \text{if dispatcher $n$ queries its server $k$ at time $t$,} \\
0, &\!\!\!\! \text{otherwise.}
\end{cases}
\label{eq:query_def}
\end{IEEEeqnarray}
Similarly, let \(a_{n,k}(t) \in \{0,1\}\) denote whether dispatcher $n$ receives an ACK/NAK from its server $k$ at time $t$, defined as
\begin{IEEEeqnarray}{rCl}
a_{n,k}(t) =
\begin{cases}
1, & \text{if dispatcher $n$ receives an ACK/NAK} \\
        & \text{from server $k$ at time $t$,} \\
0, & \text{otherwise.}
\end{cases} 
\label{eq:ack_def}
\end{IEEEeqnarray}
If dispatcher $n$ receives a status update from server \(k\) by the end of time slot \(t\), either through a query or via an ACK/NAK from job execution,  its AoI is reset to one at time slot $t+1$; otherwise, the AoI increases by one. Hence, the AoI evolves according to
\begin{IEEEeqnarray}{rCl}
\Delta_{n,k}(t+1)\! =\!
\begin{cases}
1, & \text{if } a_{n,k}(t) \!=\! 1 \text{ or } c_{n,k}(t)\! = \!1, \\
\Delta_{n,k}(t) + 1,\!\!\!\! & \text{otherwise},
\end{cases}~~~
\label{eq:aoi_update}
\end{IEEEeqnarray}
The dispatcher estimates the server's current state $X_{n,k}(t)$ using the latest observed state $X_{n,k}(t - \Delta_{n,k}(t))$ and its AoI $\Delta_{n,k}(t)$, defined as
\begin{IEEEeqnarray}{rCl}
   I_{n,k}(t) \;=\; \Bigl( X_{n,k}\bigl(t - \Delta_{n,k}(t)\bigr),\; \Delta_{n,k}(t) \Bigr).
   \label{eq:info_state}
\end{IEEEeqnarray}

Next, we evaluate the probability of successful job execution, referred to as the \emph{job success rate}. The \(\delta\)-step transition matrix of the Markov chain $X_{n,k}(t)$ is $\bm{P}_{n,k}^{\delta}$, which is the $\delta$-th power of the one-step transition matrix $\bm{P}_{n,k}$. The entries of $\bm{P}_{n,k}^{\delta}$ are denoted in the following equation \eqref{eq:transition_power}: 
\begin{IEEEeqnarray}{rCl}
\begin{bmatrix}
\phi_{n,k}(\delta) & 1 - \phi_{n,k}(\delta) \\
1 - \psi_{n,k}(\delta) & \psi_{n,k}(\delta)
\end{bmatrix}=\bm{P}_{n,k}^{\delta}.
\label{eq:transition_power}
\end{IEEEeqnarray}
If the last observed server status is $X_{n,k}(t - \Delta_{n,k}(t))=0$, then the probability that this server is idle at time $t$ is
\begin{IEEEeqnarray}{rCl}\label{eq:prob1}
\Pr \bigl[ X_{n,k}(t)=0 \mid I_{n,k}(t) \bigr] = \phi_{n,k}(\Delta_{n,k}(t)). 
\end{IEEEeqnarray}
If the last observed server status is $X_{n,k}(t - \Delta_{n,k}(t))=1$, then the probability that this server is idle at time $t$ is
\begin{IEEEeqnarray}{rCl}\label{eq:prob2}
\Pr \bigl[ X_{n,k}(t)=0 \mid I_{n,k}(t) \bigr] = 1 - \psi_{n,k}(\Delta_{n,k}(t)).
\end{IEEEeqnarray}
Each dispatcher assigns an incoming job to one of its $K$ servers with the highest probability of being idle. Therefore, if a job arrives at the dispatcher $n$ at time $t$, the probability of successful execution of this job is
\begin{align}
&r_n(t) 
=\! \max_{k \in \{1,\dots,K\}} \Pr\left[
X_{n,k}(t) = 0 \,\big|\, I_{n,k}(t)\right]. 
\label{eq:reward}
\end{align}
Hence, $r_n(t)$ is a function of $\bm{I}_{n}(t)=(I_{n,1}(t), \ldots, I_{n,K}(t))$.



\subsection{Query Scheduling}
Our objective is to develop a query scheduling policy that determines which server each dispatcher should query over time. Let $\pi =(c_{n,k}(t), t= 0,1,\ldots, n =1,\ldots,N, k = 1,\ldots,K)$ denote a query scheduling policy, and let \( \Pi \) denote the set of all causal AoI-based query scheduling policies. We assume that each dispatcher can query at most one server at any time, and that the total number of queries across all dispatchers is upper bounded by $M$ at any time $t$, with $M<N$. 
The query scheduling problem is formulated in~\eqref{eq:objective}-\eqref{eq:dispatcher_constraint}, aiming to maximize the long-term average job success rate, subject to per-time-slot channel resource constraints:
\begin{IEEEeqnarray}{cl}
\!\!\!\!\!\!\!\underset{\pi \in \Pi}{\text{sup}} \quad & 
\liminf_{T \to \infty} \frac{1}{T \sum_{n=1}^N\lambda_n} \sum_{t=0}^{T-1}   \sum_{n=1}^N
\mathbb{E}\Bigl[z_n(t) r_n(t) \Bigr]
\IEEEyesnumber\label{eq:objective} \\
\!\!\!\!\!\!\!\text{s.t.} \quad &
\sum_{n=1}^N\sum_{k=1}^K  c_{n,k}(t) \le M, \quad t = 0, 1,\ldots
\IEEEyesnumber\label{eq:channel_constraint} \\
&
\sum_{k=1}^K c_{n,k}(t) \le 1, \quad t = 0, 1,\ldots,\; n = 1,\ldots,N,
\IEEEyesnumber\label{eq:dispatcher_constraint}
\end{IEEEeqnarray}
where $z_n(t) r_n(t)$ represents the probability of successful job execution, if dispatcher $n$ has an incoming job at time $t$; it is zero otherwise. 

\section{Net-Gain Maximization Policy}
\label{sec:rmab}

The query scheduling problem~\eqref{eq:objective}-\eqref{eq:dispatcher_constraint} is a multi-action restless multi-armed bandit (RMAB) problem, where each dispatcher $n$ is treated as an arm, and its state at time $t$ is represented by the vector $\bm{I}_{n}(t)= \bigl( I_{n,1}(t), \dots, I_{n,K}(t) \bigr)$. Multi-action RMABs are PSPACE-hard \cite{papadimitriou1999complexity}. Nonetheless, we use a Net-Gain Maximization (NGM) scheduling policy~\cite{shisher2023learning, ornee2023context, chakraborty2024timely} to obtain a near-optimal solution to \eqref{eq:objective}-\eqref{eq:dispatcher_constraint}.

\subsection{Relaxation and Lagrangian Decomposition}
We relax the per-time-slot channel constraint in~\eqref{eq:channel_constraint}, substituting it with an expected time-average channel constraint, as defined in~\eqref{eq:relaxed_objective} for the relaxed problem below:
\begin{IEEEeqnarray}{cl}
\!\!\!\underset{\pi \in \Pi}{\text{sup}} \quad & 
\liminf_{T \to \infty} \frac{1}{T \sum_{n=1}^N\lambda_n} \sum_{t=0}^{T-1} \sum_{n=1}^N 
\mathbb{E}\Bigl[ z_n(t) r_n(t) \Bigr]
\IEEEyesnumber\label{eq:relaxed_objective} \\
\!\!\!\text{s.t.} \quad &
\limsup_{T \to \infty} \frac{1}{T} \sum_{t=0}^{T-1} 
\sum_{n=1}^N \sum_{k=1}^K \mathbb{E} \bigl[c_{n,k}(t)\bigr] \le M
\IEEEyesnumber\label{eq:relaxed_global_constraint} \\
&
\sum_{k=1}^K c_{n,k}(t) \le 1, \quad t = 0,1,\ldots,\; n = 1,\ldots,N.
\IEEEyesnumber\label{eq:relaxed_dispatcher_constraint}
\end{IEEEeqnarray}
Next, by introducing a Lagrange multiplier \(\mu \geq 0\) for the relaxed channel resource constraint~\eqref{eq:relaxed_global_constraint}, the Lagrangian of the problem \eqref{eq:relaxed_objective}-\eqref{eq:relaxed_dispatcher_constraint} is expressed as
\begin{IEEEeqnarray}{rCl}
\mathcal{L}(\pi, \mu) &= \liminf_{T \to \infty} \frac{1}{T}  \sum_{t=0}^{T-1} \left[ \frac{1}{\sum_{n=1}^N\lambda_n}\sum_{n=1}^N  \mathbb{E}\bigl[z_n(t)  r_n(t)\bigr] \right. \nonumber \\
&\qquad \left. - \mu \sum_{n=1}^N \sum_{k=1}^K \mathbb{E}[c_{n,k}(t)] \right] + \mu M.
\IEEEyesnumber\label{eq:lag_form}
\end{IEEEeqnarray}
The corresponding dual problem is formulated as
\begin{IEEEeqnarray}{rCl}
\mu^*=\arg\min_{\mu \geq 0} D(\mu),
\IEEEyesnumber\label{eq:dual_problem}
\end{IEEEeqnarray}
where the dual function \(D(\mu)\) is defined as
\begin{align}\label{eq:dual_function2}
\!\!\!D(\mu) = \sup_{\pi \in \Pi}~ &\mathcal{L}(\pi, \mu)\\
\text{s.t.}~ & \sum_{k=1}^K c_{n,k}(t) \le 1, t = 0,1,\ldots,n \!=\! 1,\ldots,N.\!\!
\IEEEyesnumber\label{eq:dual_function}
\end{align}
Since the term $\mu M$ in \eqref{eq:lag_form} is constant with respect to the policy \(\pi\), it can be omitted during optimization. As a result, problem \eqref{eq:dual_function} naturally separates into \(N\) independent subproblems. For a fixed \(\mu\), the subproblem for dispatcher \(n\) is defined as
\begin{IEEEeqnarray}{cl}
\!\!\!\underset{\pi_n \in \Pi_n}{\text{sup}} \quad & 
\liminf_{T \to \infty} \frac{1}{T} \sum_{t=0}^{T-1} 
\mathbb{E} \Biggl[ 
\frac{1}{\sum_{n=1}^N \lambda_n} z_n(t) r_n(t) 
\IEEEnonumber \\
& \qquad\qquad - \mu \sum_{k=1}^K c_{n,k}(t) 
\Biggr]
\IEEEyesnumber\label{eq:dispatcher_subproblem} \\
\!\!\!\text{s.t.} \quad & 
\sum_{k=1}^K c_{n,k}(t) \le 1, \quad t = 0, 1, \ldots,
\IEEEyesnumber\label{eq:dispatcher_constraint_local}
\end{IEEEeqnarray}
where \(\pi_n = (c_{n,k}(t), t= 0,1,\ldots, k=1,\ldots,K)\) denotes the query scheduling policy for dispatcher \(n\), and \(\Pi_n\) is the set of all causal AoI-based policies for dispatcher $n$.

\subsection{Optimal Primal and Dual Solutions to the Relaxed Problem}
The per-dispatcher subproblem in~\eqref{eq:dispatcher_subproblem}-\eqref{eq:dispatcher_constraint_local} is an infinite-horizon average-reward Markov Decision Process (MDP). The state of the MDP at time~$t$ is represented by the vector $\bm{I}_{n}(t)=(I_{n,1}(t), \ldots, I_{n,K}(t))$. The action at time $t$ is denoted by $A_n(t) \in \{0, 1, \dots, K\}$ where $A_n(t) = 0$ indicates that none of the $K$ servers is queried, and $A_n(t) = k\in \{ 1, 2, \dots, K\}$ indicates that server $k$ is queried at time $t$ for its updated idle/busy status.
Let $Q_{n,\mu}(\bm{I}_{n}(t), A_n(t))$ denote the relative action-value function for the MDP \eqref{eq:dispatcher_subproblem}-\eqref{eq:dispatcher_constraint_local}, which can be solved by dynamic programming.

To compute the optimal Lagrange multiplier \(\mu^*\) in \eqref{eq:dual_problem}, we apply a dual subgradient descent method. At each dual iteration \(\ell\), \(\mu_\ell\) is updated as
\begin{IEEEeqnarray}{rCl}\label{eq:dual_update}
\mu_{\ell+1} = \left[ \mu_\ell + \beta_{\ell} \left( \frac{1}{T} \sum_{t=0}^{T-1} \sum_{n=1}^N \sum_{k=1}^K c_{n,k,\mu_{\ell}}^{*}(t) - M \right)\right]_+,~~~~
\end{IEEEeqnarray}
where $[\cdot]_+=\max\{\cdot,0\}$, $\beta_l > 0$ is the step size, and $c_{n,k,\mu_{\ell}}^{*}(t)$ denotes the optimal action of the subproblem  \eqref{eq:dispatcher_subproblem}–\eqref{eq:dispatcher_constraint_local} for dispatcher $n$,
evaluated at $\mu=\mu_\ell$. One choice of the step size is $\beta_\ell = \beta/(1+\gamma\ell)$, where  $\beta$ and $\gamma$ are two parameters. This subgradient update adjusts \(\mu_\ell\) based on the deviation from the query channel resource budget \(M\). If the empirical number of queries exceeds \(M\), \(\mu_\ell\) increases to penalize over-querying; otherwise, it decreases to encourage additional queries.

After obtaining the optimal dual value \(\mu^*\), the optimal solution to the relaxed problem \eqref{eq:relaxed_objective}-\eqref{eq:relaxed_dispatcher_constraint} can be computed by substituting $\mu^*$ into \eqref{eq:dual_function2}-\eqref{eq:dual_function}, and subsequently into the decomposed subproblems \eqref{eq:dispatcher_subproblem}-\eqref{eq:dispatcher_constraint_local}. As previously mentioned, the subproblems \eqref{eq:dispatcher_subproblem}-\eqref{eq:dispatcher_constraint_local} can be solved using dynamic programming, yielding the relative action-value function $Q_{n,\mu^*}(\bm{I}_n(t), A_n(t))$. 



\subsection{Net-Gain Maximization Scheduling Policy}
\begin{algorithm}[t]
\caption{Net-Gain Maximization Scheduling}
\begin{algorithmic}[1]\label{alg1}
\STATE Input the optimal Lagrange multiplier $\mu^*$
\FOR{each time step $t = 0, 1, \dots$}
    \FOR {each dispatcher $n = 1, \dots, N$}
        \STATE Update $\bm{I}_n(t)$ based on~\eqref{eq:aoi_update}-\eqref{eq:info_state}
        \STATE Compute net-gain $\alpha_{n,\mu^*}(\bm{I}_n(t), k)$ in \eqref{eq:net_gain} for all $k$ 
        \STATE $k_n^* \leftarrow \arg\max_{k=1,\ldots,K} \alpha_{n,\mu^*}(\bm{I}_n(t), k)$
        \STATE $\alpha_n^* \leftarrow \alpha_{n,\mu^*}(\bm{I}_n(t), k_n^*)$
    \ENDFOR
    \STATE Schedule up to $M$ dispatchers with the highest non-negative gains $\alpha_n^*$
    \STATE Each scheduled dispatcher $n$ queries the $k_n^*$-th server from its group
\ENDFOR
\end{algorithmic}
\end{algorithm}

We now present a feasible solution to the original query scheduling problem~\eqref{eq:objective}–\eqref{eq:dispatcher_constraint}, based on the relative action-value function $Q_{n,\mu^*}(\bm{I}_n(t), A_n(t))$ associated with the optimal dual value $\mu^*$. This solution, termed the Net-Gain Maximization (NGM) scheduling policy~\cite{shisher2023learning,ornee2023context, chakraborty2024timely}, operates as follows: 

At the beginning of each time slot $t$, each dispatcher computes the \emph{net gain} from querying server \(k\) as
\begin{IEEEeqnarray}{rCl}
\alpha_{n,\mu^*}(\bm{I}_n(t), k) 
= Q_{n,\mu^*}(\bm{I}_n(t), k) - Q_{n,\mu^*}(\bm{I}_n(t), 0)~~~ \nonumber \\
 k \in \{1, \ldots, K\}.~~
\IEEEyesnumber\label{eq:net_gain}
\end{IEEEeqnarray}
The net gain $\alpha_{n,\mu^*}(\bm{I}_n(t), k)$ quantifies the improvement in the expected cumulative reward when dispatcher $n$ queries server $k$ at time $t$, under the relaxed problem~\eqref{eq:relaxed_objective}–\eqref{eq:relaxed_dispatcher_constraint}. The reward is defined as
the job success rate minus the dual price $\mu^*$. 
A positive net gain indicates that querying server \(k\) yields a higher expected cumulative reward than refraining from querying any server.
As summarized in Algorithm~\ref{alg1}, the NGM scheduling policy selects the servers with the highest non-negative net gains by solving the following optimization problem at each time slot~$t$: 
\begin{align}
\max_{c_{n,k}(t), \forall k, \forall n} ~& \sum_{n=1}^N\sum_{k=1}^K \alpha_{n,\mu^*}(\bm{I}_n(t), k) c_{n,k}(t) \\
\text{s.t.} ~~~~~~&\sum_{n=1}^N\sum_{k=1}^K c_{n,k}(t)\leq M, \\
~~~~~~&\sum_{k=1}^K c_{n,k}(t)\leq 1, ~n=1,\ldots,N.
\end{align}



\begin{figure}[t]
    \centering
    \includegraphics[width=0.8\linewidth]{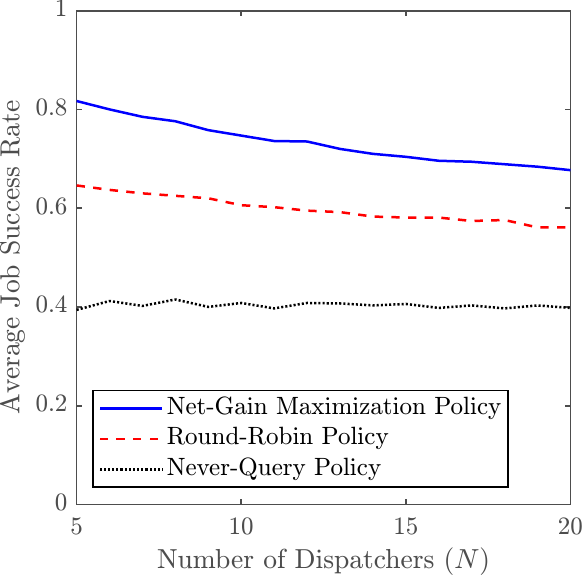}
    \caption{Average job success rate with increasing number of dispatchers ($N$), 
    where the number of channels is $M=5$, the number of servers per dispatcher is $K=5$, 
    and the job arrival probability is $\lambda_n=0.3$.}
    \label{fig:avg_reward_vs_N}
\end{figure}

\begin{figure}[t]
    \centering
    \includegraphics[width=0.8\linewidth]{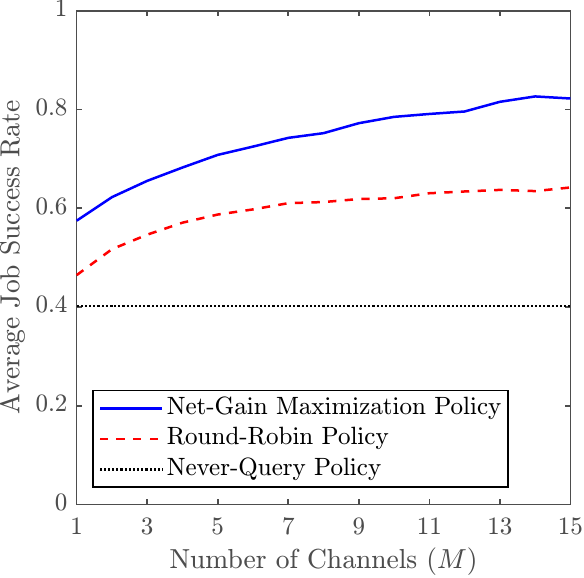}
    \caption{Average job success rate with increasing number of channels ($M$), 
    where the number of dispatchers is $N=15$, the number of servers per dispatcher is $K=5$, and the job arrival probability is $\lambda_n=0.3$.}
    \label{fig:avg_reward_vs_M}
\end{figure}
\section{Numerical Results}
To assess the scalability of our system, we compare three query allocation policies: (i) Never-Query Policy, which does not issue explicit queries and relies solely on job execution feedback; (ii) Round-Robin Policy, which queries servers in a uniform cyclic order without adapting to server states; and (iii) our proposed NGM Policy. Similar to our model, in the baseline policies each dispatcher assigns arriving jobs to the server with the highest estimated idle probability, and this estimate is continuously refreshed based on the implicit feedback (ACK/NAK) from job assignments. In these evaluations, we set $\phi_{n,k}=0.85$ and $\psi_{n,k}=0.90$ for all dispatchers $n$ and servers $k$, each dispatcher has an arrival probability $\lambda_n=0.3$, and the maximum AoI is capped at 20. We then systematically evaluate the impact of two key factors: the number of dispatchers ($N$) and the number of channels ($M$).



Figure~\ref{fig:avg_reward_vs_N} shows the impact of the number of dispatchers on the average job success rate when the number of channels is fixed at $M=5$ and each dispatcher manages $K=5$ servers. The results indicate that the job success rate gradually decreases as $N$ increases from 5 to 20, since a fixed number of channels must be distributed across more dispatchers, thereby reducing querying opportunities per dispatcher. Across all cases, the NGM Policy achieves the highest performance, with up to a 26\% gain over the Round-Robin Policy and up to a 107\% gain over the Never-Query Policy. These findings demonstrate that NGM is able to maintain superior execution performance even as query resources per dispatcher diminish with increasing system scale.

Figure~\ref{fig:avg_reward_vs_M} shows the impact of the number of channels on the average job success rate when the number of dispatchers is fixed at $N=15$ and each dispatcher manages $K=5$ servers. The results indicate that the job success rate increases as $M$ grows, since additional channels provide more querying opportunities and improve the freshness of server state information. Among the evaluated policies, NGM consistently achieves the best performance, with up to a 30\% gain over the Round-Robin Policy and up to a 105\% gain over the Never-Query Policy. These findings confirm that NGM effectively leverages additional communication resources to deliver improved job execution performance.

\section{Conclusion}
In this paper, we studied the goal-oriented communication problem of server availability monitoring for real-time job assignment in edge AI. We introduced a Net-Gain Maximization scheduling algorithm to query the availability statuses of edge servers and maximize the success rate of AI compute tasks. 
Our results show that the NGM Policy significantly improves server availability awareness and job-success rates compared to baseline policies, achieving up to a 30\% gain over the Round-Robin Policy and up to a 107\% gain over the Never-Query Policy. As an extension to this work, we plan to incorporate task buffering and failover mechanisms into the framework to further enhance system performance.

\bibliographystyle{IEEEtran}
\bibliography{references}

\end{document}